\documentclass[twocolumn,aps,pra,showpacs,amsmath,superscriptaddress,
amsfonts,amssymb,floatfix]{revtex4}
\usepackage{booktabs}
\usepackage{graphicx}
\usepackage{mathrsfs}
\usepackage{epsfig}
\usepackage{graphicx}
\usepackage{amsmath}

\begin{document}

\title{Optimized Dynamical Decoupling Sequences in Protecting Two-Qubit States}
\author{Yu Pan}
\affiliation{Department of Physics, National University of
Singapore, 117542, Republic of Singapore}
\affiliation{Key
Laboratory of Systems and Control, Institute of Systems Science,
Academy of Mathematics and Systems Science, Chinese Academy of
Sciences, Beijing 100190, People's Republic of China}
\author{Zai-Rong Xi}
\affiliation{Key
Laboratory of Systems and Control, Institute of Systems Science,
Academy of Mathematics and Systems Science, Chinese Academy of
Sciences, Beijing 100190, People's Republic of China}
\author{Jiangbin Gong}
\email{phygj@nus.edu.sg}
\affiliation{Department of Physics, National University of
Singapore, 117542, Republic of Singapore}
\affiliation{Centre
for Computational Science and Engineering, National University of
Singapore, 117542, Republic of Singapore}
\affiliation{NUS Graduate School for Integrative
Sciences and Engineering, 117597,
Republic of Singapore}

\begin{abstract}
Aperiodic dynamical decoupling (DD) sequences of $\pi$ pulses are of great interest
to decoherence control and have been recently
extended from single-qubit to two-qubit systems. If
the environmental noise power spectrum is made available, then one may further optimize
aperiodic DD sequences
to reach higher efficiency of decoherence suppression than known universal schemes.
This possibility is
investigated in this work for the protection of two-qubit
states, using an exactly solvable pure dephasing model
including both local and nonlocal noise.  The performance of optimized DD sequences in protecting two-qubit
states is compared with that achieved by nested Uhrig's DD (nested-UDD) sequences, for several different types of noise spectrum.
Except for cases with noise spectrum decaying slowly in the high-frequency regime,
optimized DD sequences with tens of control pulses can
perform orders of magnitude better than that of nested-UDD.
A two-qubit system with highly unbalanced local noise is also examined to shed more light on
a recent experiment. Possible experiments that may be motivated by this work are discussed.
\end{abstract}

\date{\today}

\maketitle

\section{INTRODUCTION}
In one way or another any two-level system (qubit)  is coupled to some environmental degrees of freedom.
This inevitable system-environment coupling leads to decoherence. To protect quantum coherence as a key
resource for quantum technologies, many schemes have been proposed to
dynamically eliminate the unwanted qubit-environment coupling~\cite{ar1,ar5,ar2,ar3,ar3n}.

Analogous to the spin-echo technique widely adopted in
nuclear-magnetic-resonance studies~\cite{ar4}, various dynamical
decoupling (DD) sequences of instantaneous control
pulses~\cite{ar5,ar6,ar7,ar8} have been extensively studied. In
particular, due to Uhrig's DD (UDD)  sequence~\cite{ar7,ar8},
research activities focusing on DD have surged
recently~\cite{ar9,ar14,ar15,ar17,ar18,ar19}. In addition to its
high efficiency in theory, the power of UDD lies in its
universality~\cite{ar10,arreview}. Indeed, the working mechanism of
UDD does not rely on detailed assumptions about the
system-environment coupling or about the environment. Nevertheless,
if the actual form of the noise spectrum of the environment is
available, then a locally optimized DD (LODD) \cite{ar9} sequence
based on the noise spectrum can outperform UDD, with the pulse
locations optimized according to an exact decoherence
function~\cite{ar11,ar12,ar13}. The reason for the success of
optimization is simple. In suppressing the pure-dephasing of a
qubit, a UDD sequence minimizes a decoherence filter function
(defined later) in the neighborhood of zero frequency but gradually
becomes less effective for large frequencies. As such, if the noise
spectrum of the environment is known, then its actual behavior at
appreciably nonzero frequencies makes room for further optimization
of DD sequences. This optimization approach is somewhat in the same
spirit of the continuous DD approach ~\cite{ar2,ar3}, insofar as
both attempt to make the full use of the noise spectrum.

Extension of DD to two-qubit (or even multi-qubit) systems is
crucial towards efficient protection of quantum entanglement. For a
known initial state of a two-qubit system, an extended UDD sequence
is found by involving nonlocal control operators, with its
theoretical performance essentially identical with that in one-qubit
UDD cases~\cite{ar20}. For general situations, nested-UDD sequences,
initially proposed for suppressing both dephasing and relaxation in
one-qubit systems~\cite{ar21}, are advocated to protect two-qubit
states in a universal manner, i.e., without knowing the details of
the system-environment coupling or of the noise
spectrum~\cite{ar22,ar23}.   For example, it was shown that to lock
an unknown superposition state of two known basis states to the
$N$th order, three layers of UDD sequences and hence about $N^3$
control pulses in total will be needed~\cite{ar20}.  If the initial
state is totally unknown, then to reach the same level of
decoherence suppression one needs four layers of UDD sequences and
hence about $N^4$ control pulses~\cite{ar22,ar23}. One important
question then arises.  That is, if the noise spectrum of the
environment of a two-qubit system is available, then can we
significantly improve entanglement protection by further optimizing
the locations of the instantaneous control pulses as what was done in
\cite{ar9,ar11} for single-qubit systems? If yes, then the required number
of pulses can be much less and more understandings of entanglement
protection might emerge.  Using an approach extended from the
above-mentioned LODD for single-qubit systems, this question is
answered here via a pure dephasing model of an open two-qubit
system. The performance of optimized DD sequences in preserving
two-qubit states is compared with that achieved by nested-UDD, for
several different types of noise spectrum.   Except for the
Lorentzian type of noise spectrum, it is found that optimized DD
sequences can protect two-qubit states orders of magnitude better
than nested-UDD.  As a result, on the one hand nested-UDD can be
seen as a powerful and a general-purpose scheme for protecting
two-qubit states, and on the other hand the optimized DD approach
can be seen as system-specific DD schemes with even better
performance. In addition, to shed more light on a recent experiment
of entanglement protection via DD~\cite{ar24}, a two-qubit system
that shares important noise features with the experiment is studied.

Our plan for this paper is as follows. In the next section, we
discuss a model that describes pure dephasing processes of a  two-qubit
system in the presence of instantaneous $\pi$ pulses. Based on exact expressions of decoherence effects,
we elaborate in Sec. III our optimization procedure to find the optimized pulse locations.
In Sec. IV, the performance of optimized two-qubit DD sequence
is illustrated in four subsections treating different types of noise spectrum.
Sec. V concludes this work and proposes two types of experiments.

\section{Pure Dephasing Model of Two-Qubit Systems}
In general, decoherence as a complicated process involves both
population relaxation and dephasing.  Yet, if we increase the
strength of an external polarization field such that all the energy
level splittings are sufficiently large, then the relaxation can be
made negligible within a time scale of interest and dephasing
becomes the only source of decoherence. Under such a pure dephasing
assumption, the environmental noise may be modeled as classical
random fields causing random phase shifts, a valid treatment in many
dephasing environments like spin bath in solid-state systems
\cite{ar9,ar15,ar17} and background noise in superconducting qubits
\cite{ar25}. Here we follow the methodology proposed in
\cite{ar8,ar25}. We can then write the pure-dephasing Hamiltonian of
an open two-qubit system as
\begin{eqnarray}
H=f_1(t)\sigma_{z_1}+f_2(t)\sigma_{z_2}+f_3(t)\sigma_{z_1}\sigma_{z_2},
\end{eqnarray}
where $f_i(t)$ are assumed to be independent random noise variables with a Gaussian distribution.  We further assume
\begin{eqnarray}
&\langle{f_i}\rangle=0,&\\
&\langle{f_i(t_1)f_i(t_2)\rangle=g_i(t_1-t_2)},&
\end{eqnarray}
with the correlation function $g_i(t)$ being an even function.  The $f_1(t)$ [$f_2(t)$] term in the Hamiltonian
depicts the noise seen by the first (second) qubit alone;
whereas the $f_3(t)$ term reflects how noise may jointly impact the two qubits. In the following we loosely call
the $f_3(t)$ noise term as a term of nonlocal noise.  If the two qubits are far apart, then this nonlocal noise
term should be very small and can be neglected. As a consequence the decoherence control problem is expected to share important features
with single-qubit DD~\cite{agarwal} (note that $f_1(t)$ and $f_2(t)$ are assumed to be independent here).
However, of our concern in this study are two nearby qubits interacting with each other, and therefore the nonlocal noise term should be included to account for random fluctuations in the qubit-qubit mutual interaction (for example, due to the fluctuations
in qubit-qubit distance caused by lattice vibrations, or in the context of super-conducting qubits~\cite{ar27n}, due to the fluctuations in a third device
 that is responsible for qubit-qubit coupling).  Certainly, in a more realistic environment the $f_3(t)$ term might be correlated with the local noise terms. Such kind of
potential correlations are neglected in our model. From a different
perspective, one may regard the first qubit as a part of the environment of the second qubit, and then the $f_3(t)$ term models how the dynamics of one qubit might change the environment experienced by the other qubit.   Throughout we assume dimensionless units.

Let $|\uparrow\rangle$  and $|\downarrow\rangle$ be the eigenstates of $\sigma_z$, then
a general two-qubit pure state at time zero can be written as
\begin{eqnarray}
|\Psi(0)\rangle
=\alpha|\downarrow\downarrow\rangle
+\beta|\downarrow\uparrow\rangle+\gamma|\uparrow\downarrow\rangle+\eta|\uparrow\uparrow\rangle,
\end{eqnarray}
where the upward or downward arrows in each of the four components represent the spin states of the two qubits.
The Hamiltonian in Eq.~(1) then gives rise to the following state vector at time $t$,
\begin{eqnarray}
|\Psi(t)\rangle&=&\alpha{e^{-\mbox{i}[-F_1(t)-F_2(t)+F_3(t)]}}|\downarrow\downarrow\rangle\nonumber\\
&+&\beta{e^{-\mbox{i}[-F_1(t)+F_2(t)-F_3(t)]}}|\downarrow\uparrow\rangle\nonumber\\
&+&\gamma{e^{-\mbox{i}[F_1(t)-F_2(t)-F_3(t)]}}|\uparrow\downarrow\rangle\nonumber\\
&+&\eta{e^{-\mbox{i}[F_1(t)+F_2(t)+F_3(t)]}}|\uparrow\uparrow\rangle,
\end{eqnarray}
where $F_i(t)\equiv \int_0^tf_i(t^{'})dt^{'}$.  To analyze the density matrix $\rho(t)=|\Psi(t)\rangle\langle\Psi(t)|$ averaged over noise histories (denoted $\langle\cdot\rangle$), we further define
four basis states $|0\rangle=|\downarrow\downarrow\rangle,
|1\rangle=|\downarrow\uparrow\rangle,
|2\rangle=|\uparrow\downarrow\rangle,
|3\rangle=|\uparrow\uparrow\rangle.$  Then all the averaged density matrix elements (in the absence of DD control pulses) can be easily worked out.
For example, the mean value of $\rho_{01}(t)$, denoted $\bar{\rho}_{01}(t)$,
can be expressed as
\begin{eqnarray}
\bar{\rho}_{01}(t)&=&\alpha^\ast\beta{\langle}e^{-\mbox{i}[F_2(t)-F_3(t)]}e^{-\mbox{i}[F_2(t)-F_3(t)]}\rangle\nonumber\\
&=&\alpha^\ast{\beta}e^{-2\langle{[F_2(t)-F_3(t)]}^2\rangle}\nonumber\\
&=&\alpha^\ast{\beta}e^{-2[\langle F^2_2(t)\rangle+\langle F_3^2(t)\rangle]}.
\end{eqnarray}
Here, in obtaining the last equality we have used the relation
${\langle}f_2f_3\rangle=0$ as well as the Gaussian nature of the noise.
Similar expressions can be obtained for all other density matrix elements.

In our pure-dephasing model, the error operator for the first
(second) qubit is $\sigma_{z_1}$ ($\sigma_{z_2}$) only, whose
detrimental effect can be suppressed by a local control operator
$\sigma_{x_1}$ ($\sigma_{x_2}$).  As seen from Refs.
\cite{ar22,ar23}, a nested-UDD scheme with two layers of
$\sigma_{x_1}$ and $\sigma_{x_2}$ pulses can suppress the dephasing
to the $N$th order, with about $N^2$ pulses in total. To examine if
we can further improve the performance by optimizing the pulse
locations, let us now consider the following scenario: $n$ pulses of
$\pi$ rotation along axis-$x$ are applied to the first qubit, with
the pulse locations given by $t_1, t_2,..., t_n$, whereas $m$ analogous pulses
are applied to the second qubit at times $t_{1^{'}}, t_{2^{'}},...,
t_{m^{'}}.$   As a result, totally $n+m$ pulses are applied to the
two-qubit system. We arrange the $n+m$ pulse timings in increasing
order and denote them by  $t_{1^{''}}, t_{2^{''}},...,
t_{{(n+m)}^{''}}$, with $t_{j^{''}}<t_{(j+1)^{''}}$. At each of such
instants, either the $\sigma_{z_1}$ or $\sigma_{z_2}$ operator
switches its sign. At the same time, the operator
$\sigma_{z_1}\sigma_{z_2}$ changes its sign $n+m$ times at these
instants.  This motivates us to define three switch functions~\cite{ar8}:
\begin{eqnarray*}
s_1(t^{'})=(-1)^{k_1}, & t_{k_1}<t^{'}{\le}t_{(k_1+1)}, & k_1=0,1,..., n,\\
s_2(t^{'})=(-1)^{k_2}, & t_{k_2^{'}}<t^{'}{\le}t_{(k_2+1)^{'}}, & k_2=0,1,..., m,\\
s_3(t^{'})=(-1)^{k_3}, & t_{k_3^{''}}<t^{'}{\le}t_{(k_3+1)^{''}}, &
k_3=0,1,..., n+m,
\end{eqnarray*}
with $t_0=0$, and $ t_{n+1}=t_{{(m+1)}^{'}}=t_{(n+m+1)^{''}}=t$.
For times outside the domain $[0,t]$ these switch functions are defined to be zero.
The influence of the DD pulses can then be expressed in a rather compact form.
Still taking the decay of $\bar{\rho}_{01}$ as an example, we obtain
\begin{eqnarray}
\bar{\rho}_{01}(t)=\alpha^\ast{\beta}e^{-2[\langle{ \tilde{F}^2_2(t) \rangle+\langle \tilde{F}^2_3(t)}\rangle]},
\label{decay}
\end{eqnarray} where
$\tilde{F}_i(t)$ is given by
\begin{eqnarray}
\tilde{F}_i(t)=\int_{-\infty}^{\infty}f_i(t^{'})s_i(t^{'})dt^{'}.
\end{eqnarray}

To proceed we next define three filter functions~\cite{ar8} from the Fourier
transform of $s_i(t)$, i.e.,
\begin{eqnarray}
\int_{-\infty}^{\infty}s_1(t^{'})e^{\mbox{i}\omega{t^{'}}}dt^{'}&=&\frac{\mbox{i}}{\omega}y_n(\omega{t}),\\
\int_{-\infty}^{\infty}s_2(t^{'})e^{\mbox{i}\omega{t^{'}}}dt^{'}&=&\frac{\mbox{i}}{\omega}y_m(\omega{t}),\\
\int_{-\infty}^{\infty}s_3(t^{'})e^{\mbox{i}\omega{t^{'}}}dt^{'}&=&\frac{\mbox{i}}{\omega}y_{n+m}(\omega{t}).
\end{eqnarray}
Here, the filter function with $M$ (which is $n$, $m$, or $n+m$) pulses
is defined as
\begin{eqnarray}
y_M(\omega{t})=1+(-1)^{M+1}e^{\mbox{i}\omega{t}}+2\sum_{j=1}^M(-1)^je^{\mbox{i}\omega{t}\delta_j},
\end{eqnarray} with
$\delta_j$ being the pulse location normalized to the total duration
$t$, i.e., $\delta_j=t_j/t$.  The decay factor in the exponential of Eq.~(\ref{decay})
then reduces to
\begin{eqnarray}
&&\langle{\tilde{F}^2_2(t)\rangle+\langle\tilde{F}_3^2(t)}\rangle\nonumber\\
&=&\int_{-\infty}^{\infty}\int_{-\infty}^{\infty}dt_1dt_2\ s_2(t_1)g_2(t_1-t_2)s_2(t_2) \nonumber \\
&&+\int_{-\infty}^{\infty}\int_{-\infty}^{\infty}dt_1dt_2\ s_3(t_1)g_3(t_1-t_2)s_3(t_2)\nonumber\\
&=&\frac{1}{\pi}\int_0^{\infty}|y_m(\omega{t})|^2\frac{S_2(\omega)}{\omega^2}d\omega \nonumber \\
&&+
\frac{1}{\pi}\int_0^{\infty}|y_{m+n}(\omega{t})|^2\frac{S_3(\omega)}{\omega^2}d\omega,
\\  \nonumber
\end{eqnarray}
where $S_i(\omega)$ is the noise spectrum (or spectral density), namely, the Fourier transform of the noise
correlation
function $g_i(t)$, with
\begin{eqnarray}
g_i(t)=\frac{1}{\pi}\int_0^{\infty}S_i(\omega)\cos(\omega{t})d\omega.
\\ \nonumber
\end{eqnarray}
Note that an equivalent expression may be derived using an exactly
solvable spin-boson model with pure-dephasing. Interestingly, the
decay factor in Eq.~(\ref{decay}) is seen to be the sum of two
decoherence functions, each being analogous to that in the previous
DD studies~\cite{ar7,ar8}. Yet, the problem here is still rather
complicated for two reasons. First, for the same control pulse, the
coefficient $(-1)^{j}$ can take different values in the three filter
functions. Indeed, its actual value depends on its relative pulse
location in the time sequences of $t_{k_{1}}$, of $t_{k_{2}^{'}}$,
or of $t_{k_{3}^{''}}$.  Second, density matrix elements do not
decay in the same fashion, and hence the decay of all off-diagonal
density matrix elements needs to be accounted for.

Repeating the above procedure for all other density matrix elements, we finally obtain
the averaged full density matrix $\bar{\rho}(t)$ as follows:
\begin{widetext}
$$\left(
  \begin{array}{cccc}
    |\alpha|^2 & \alpha^\ast\beta{e^{-\Gamma_2(t)-\Gamma_3(t)}} & \alpha^\ast\gamma{e^{-\Gamma_1(t)-\Gamma_3(t)}} & \alpha^\ast\eta{e^{-\Gamma_2(t)-\Gamma_1(t)}} \\
    \beta^\ast\alpha{e^{-\Gamma_2(t)-\Gamma_3(t)}} & |\beta|^2 & \beta^\ast\gamma{e^{-\Gamma_2(t)-\Gamma_1(t)}} & \beta^\ast\eta{e^{-\Gamma_1(t)-\Gamma_3(t)}}\\
    \gamma^\ast\alpha{e^{-\Gamma_1(t)-\Gamma_3(t)}} & \gamma^\ast\beta{e^{-\Gamma_2(t)-\Gamma_1(t)}} & |\gamma|^2 & \gamma^\ast\eta{e^{-\Gamma_2(t)-\Gamma_3(t)}}\\
    \eta^\ast\alpha{e^{-\Gamma_2(t)-\Gamma_1(t)}} & \eta^\ast\beta{e^{-\Gamma_1(t)-\Gamma_3(t)}} & \eta^\ast\gamma{e^{-\Gamma_2(t)-\Gamma_3(t)}} & |\eta|^2
  \end{array}
\right),$$
\end{widetext}
with the decay exponents for the off-diagonal elements given by
\begin{eqnarray}
\Gamma_1&=&\int_0^\infty|y_n|^2\frac{S_1(\omega)}{\omega^2}d\omega, \label{shortform1}\\
\Gamma_2&=&\int_0^\infty|y_m|^2\frac{S_2(\omega)}{\omega^2}d\omega,\label{shortform2}\\
\Gamma_3&=&\int_0^\infty|y_{n+m}|^2\frac{S_3(\omega)}{\omega^2}d\omega.
\label{shortform3}
\end{eqnarray}
For convenience, unimportant constants in
Eqs.~(\ref{shortform1})-(\ref{shortform3}) are either set to unity
(e.g., $t=1$) or absorbed into the noise spectrum.

Before ending this section, we stress that our straightforward calculations
above are made possible by first assuming the statistical
independence of $f_1(t)$ and $f_2(t)$.  If $f_1(t)$ and $f_2(t)$ has nonzero
correlations (noise under this correlated situation may be also called nonlocal noise, which is
much different from our case here),
then many of the density matrix elements cannot be evaluated analytically.

\section{Optimization Procedure}
Next we aim to optimize the pulse locations to keep $\bar{\rho}(t)$ as
close as possible to the initial state. Taking the trace fidelity
$C(t)=\text{Tr}[\bar{\rho}(t)\rho(0)]$ as a measure of DD performance, it is
straightforward to carry out the optimization if the initial state is known.
However, in many cases a two-qubit state to be protected or stored is unknown,
and as such the average fidelity for all possible initial states can be of more interest.
Averaging $C(t)$ over all initially
pure states and using the fact that
$\langle{|\alpha|^2}\rangle=\langle{|\beta|^2}\rangle=\langle{|\gamma|^2}\rangle=\langle{|\eta|^2}\rangle=\frac{1}{4}$,
we arrive
at
\begin{equation}
\bar{C}(t)=\frac{1}{4}+\frac{1}{4}(e^{-\Gamma_1-\Gamma_2}+e^{-\Gamma_1-\Gamma_3}+e^{-\Gamma_2-\Gamma_3}).
\end{equation}
The optimization then becomes
the minimization of $\Phi(t)\equiv 4[1-C(t)]$.  The function $\Phi(t)$, called the performance function below, is given by
\begin{eqnarray}
\Phi(t)=3-(e^{-\Gamma_1-\Gamma_2}+e^{-\Gamma_1-\Gamma_3}+e^{-\Gamma_2-\Gamma_3}).
\end{eqnarray}
Clearly $0\leq \Phi(t) \leq 3$, and a smaller value of $\Phi(t)$ indicates a higher degree of decoherence suppression.
Therefore $\Phi(t)$ can be also regarded an error function.

To minimize the value of the performance function, we first consider a total of $n+m$ pulses, among which
$m$ pulses are applied to the second qubit. Assuming that the $n+m$ pulses are applied at the timings $$\delta_1<\delta_2<\delta_3<...<\delta_{n+m},$$  one first needs to pick out $m$ pulse locations to apply the $\sigma_{x_{2}}$ control operator.
The number of such choices is given by the combination $C^{m}_{n+m}$.  Then the analytic form of the three filter functions $y_n(\omega)$, $y_m(\omega)$, and $y_{n+m}(\omega)$ can be fixed.  We next minimize the performance function $\Phi(t)$
by optimizing the $(n+m)$ pulse locations using the line-search algorithm.  In every iteration, the algorithm searches for the minimum along the direction set by a gradient. We use $n+m$ equally spacing pulses and a two-layer nested-UDD
sequence as our initial guesses.  Once the result converges, we obtain the
locally optimized locations for $n+m$ control pulses in a particular
$(n+m,m)$ scheme. To find the optimal pulse locations, we test all $C_{n+m}^m$ possible choices of pulse allocations for the second qubit.  Furthermore, by scanning the value of $m$
from $1$ to the total number of pulses, we can find an optimal pulse-number partition for a fixed total number of pulses.

\section{Performance of Optimized Sequences}
\begin{table}
\caption{Impact of nonlocal noise on the performance of optimized
two-qubit DD sequences. The local noise spectrum is assume to be
$S_1(\omega)=\omega\Theta(1-\omega)$ for the first qubit and
$S_2(\omega)=\omega\Theta(1-\omega)$ for the second qubit.
$n+m=8$. In the column of pulse location, a listed
integer $j$ means that the $j$th pulse will be a $\sigma_{x_2}$ pulse (i.e., applied to
the second qubit). Performance
is optimized by considering all $C_8^2$ and $C_8^4$
possible pulse allocations for the two qubits. \label{ta1}}
\begin{ruledtabular}
\begin{tabular}{ccc}
Nonlocal spectrum $S_3(\omega)$ & Pulse location & Performance \\
$2\omega\Theta(2-\omega)$ & 2,4,6,8 & $4.59\times10^{-5}$\\
$0.5\omega\Theta(0.5-\omega)$ & 2,4,6,8 & $4.59\times10^{-5}$\\
$0.1\omega\Theta(0.1-\omega)$ & 2,4,6,8 & $4.43\times10^{-5}$\\
$\frac{0.2}{\omega^2+1}$(infinite cutoff) & 2,4,6,8 & $1.67\times10^{-3}$\\
no nonlocal noise & 2,4,6,8 & $4.08\times10^{-10}$
\end{tabular}
\end{ruledtabular}
\end{table}
In the
following, for convenience we set $t=1$ when comparing cases of different noise
spectrum. Effects of varying $t$ can be understood as the result of
a rescaling of the strength and shape of the noise spectrum~\cite{ar13}. To appreciate that the issue of two-qubit
decoherence suppression is significantly different from a one-qubit
problem, let us first illustrate the influence of the nonlocal noise
term $f_3(t) \sigma_{z_1}\sigma_{z_2}$ on the performance of DD with
$n+m=8$. Computational examples are shown in Table~\ref{ta1}. It is
seen that in the absence of nonlocal noise, the optimized
performance (error) is of the order of $10^{-10}$, and the optimized
DD pulses are applied at
$$[0.10\quad0.10\quad0.35\quad0.35\quad0.65\quad0.65\quad0.90\quad0.90],$$
which is simply two optimized 4-pulse sequences simultaneously applied to the two qubits.
However, this DD sequence cannot suppress any nonlocal noise because
every pair of such simultaneous pulses will keep the sign of $\sigma_{z_1}\sigma_{z_2}$ unchanged, and the associated
decay exponent $\Gamma_3$ would be the same as that without DD, which is given by
\begin{eqnarray}
\Gamma_3=4\int_0^{\infty}S_3(\omega)\frac{\sin^2(\omega/2)}{\omega^2}d\omega.
\end{eqnarray}
Interestingly, as shown in Table~\ref{ta1}, even with a very weak
nonlocal noise added, e.g.,
$S_3(\omega)=0.1\omega\Theta(0.1-\omega)$, where $\Theta(\cdot)$ is the Heaviside step function, the minimal error that can be achieved by an optimized DD sequence
is already increased by
 five orders of magnitude!  This clearly
addresses the importance of taking nonlocal noise into
consideration for decoherence suppression.  With this understanding we are now ready to examine the
optimization of DD in two-qubit systems.

\subsection{Spectrum with Hard Cutoff}
\begin{table}
\caption{Performance of optimized two-qubit DD sequences as compared with nested-UDD, for
a few examples of local and nonlocal noise spectrum. Performance is optimized
for each combination $C_{n+m}^{m}$ taking into account all possibilities of
allocating $m$ pulses to the second qubit.  In the column of pulse location,
a listed integer $j$ means that the $j$th pulse is a $\sigma_{x_2}$ pulse (i.e., applied the second qubit)
during the first half of the dynamics. The second half of the pulse sequence
is symmetric to the first half. Ohmic noise spectrum with hard cutoff is considered. \label{ta2}}
\begin{ruledtabular}
\begin{tabular}{ccc}
\multicolumn{3}{c}{$S_1=\omega\Theta(1-\omega),S_2=\omega\Theta(1-\omega),S_3=2\omega\Theta(2-\omega)$}\\
\hline
Pulse combination & Pulse location & Performance \\
\hline
nested-UDD(2) & 3 & $7.32\times10^{-4}$\\
$C_8^2$ & 3 & $8.66\times10^{-5}$\\
$C_8^4$ & 2,4 & $4.59\times10^{-5}$\\
nested-UDD(3) & 4,8 & $2.45\times10^{-6}$\\
$C_{15}^3$ & 4,8 & $3.04\times10^{-7}$\\
$C_{15}^5$ & 2,5,8 & $6.14\times10^{-9}$\\
$C_{15}^9$ & 1,3,5,7,8 & $1.17\times10^{-10}$\\
\hline
\hline
\multicolumn{3}{c}{$S_1=\omega\Theta(1-\omega),S_2=\omega\Theta(1-\omega),S_3=0.5\omega\Theta(0.5-\omega)$}\\
\hline
Pulse combination & Pulse location & Performance \\
\hline
nested-UDD(2) & 3 & $3.26\times10^{-4}$\\
$C_8^2$ & 3 & $8.14\times10^{-5}$\\
$C_8^4$ & 2,4 & $4.59\times10^{-5}$\\
nested-UDD(3) & 4,8 & $1.66\times10^{-6}$\\
$C_{15}^3$ & 3,8 & $1.88\times10^{-7}$\\
$C_{15}^5$ & 3,5,8 & $7.06\times10^{-11}$\\
$C_{15}^9$ & 1,3,4,6,8 & $6.26\times10^{-10}$
\end{tabular}
\end{ruledtabular}
\end{table}
Previous work showed that in single-qubit cases, an
optimized DD sequence can greatly outperform UDD for an Ohmic
noise spectrum with hard cutoff \cite{ar9,ar11,ar12,ar13}. It is observed that
this finding also holds in our two-qubit dephasing model. Here, we set the local noise
spectrum of the two qubits to be the same, and then vary the intensity of the
nonlocal noise spectrum, from twice as much as the local noise spectrum to a relatively weak
one.  As seen from Table~\ref{ta2}, the optimized DD sequence is in general
much better than nested-UDD. In particular, for
a 15-pulse sequence ($n+m=15$), the optimization improves the performance by many orders of
magnitude as compared with a two-layer nested-UDD(3)
(here nested-UDD($k$) means that $k$ pulses applied to the second qubit in the outside layer, and
 within each layer a $k$-pulse UDD sequence is applied to the first qubit).
 As a matter of fact, by comparing Table~\ref{ta2} and Table~\ref{ta6}, it is seen
 that the best performance of optimized 15-pulse DD sequence for
$S_3=0.5\omega\Theta(0.5-\omega)$ can even be two orders of magnitude better
than nested-UDD(4), which requires 24 pulses.
This indicates that
if the noise spectrum is known, then the required pulse number to achieve a given degree of decoherence suppression
does not have to scale with
$N^2$ as in nested-UDD($N$).  Further, as suggested by the results in Table~\ref{ta2}, to achieve the best performance
the number of $\sigma_{x_2}$ pulses
is larger than that suggested by nested-UDD. But interestingly, this does not necessarily mean that the number of $\sigma_{x_2}$ pulses should be as close as possible to the number of $\sigma_{x_1}$ pulses. For example,  the $C^5_{15}$ (instead of $C^7_{15}$)
case shown in the bottom of Table II gives the best performance.
This is one of the subtle consequences of nonlocal noise spectrum.  Lastly, as $S_3(\omega)$ changes from $2\omega\Theta(2-\omega)$ to $0.5 \omega\Theta(0.5-\omega)$, it is seen from Table II that not only the optimized performance changes, but also the whereabouts of $\sigma_{x_2}$ pulses become much different.  The physical intuition why the $\sigma_{x_2}$ pulses should be shifted in this manner is far from obvious, reflecting the importance of the relative locations of $\sigma_{x_2}$ pulses with respect to $\sigma_{x_1}$ pulses.

\begin{table}
\caption{Same as in Table II, but for Ohmic
noise spectrum with  hard cutoff at larger frequencies and for $1/f$ spectrum with hard cutoff. \label{ta3}}
\begin{ruledtabular}
\begin{tabular}{ccc}
\multicolumn{3}{c}{$S_1=\omega\Theta(5-\omega),S_2=\omega\Theta(5-\omega),S_3=\omega\Theta(3-\omega)$}\\
\hline
Pulse combination & Pulse location & Performance \\
\hline
nested-UDD(2) & 3 & $1.55$\\
$C_8^2$ & 3 & $0.80$\\
$C_8^4$ & 2,4 & $0.54$\\
nested-UDD(3) & 4,8 & $0.36$\\
$C_{15}^3$ & 3,8 & $6.63\times10^{-2}$\\
$C_{15}^7$ & 2,4,6,8 & $1.48\times10^{-6}$\\
\hline
\hline
\multicolumn{3}{c}{$S_1=\frac{1}{\omega}\Theta(10-\omega),S_2=\frac{1}{\omega}\Theta(10-\omega),S_3=\frac{1}{\omega}\Theta(5-\omega)$}\\
\hline
Pulse combination & Pulse location & Performance \\
\hline
nested-UDD(2) & 3 & $0.61$\\
$C_8^2$ & 3 & $0.60$\\
$C_8^4$ & 2,4 & $0.41$\\
nested-UDD(3) & 4,8 & $0.32$\\
$C_{15}^3$ & 4,8 & $0.22$\\
$C_{15}^7$ & 2,4,6,8 & $9.96\times10^{-5}$
\end{tabular}
\end{ruledtabular}
\end{table}

In single-qubit cases~\cite{ar8,ar11,ar12,ar13}, LODD is more
powerful for a noise spectrum with a larger cutoff. This motivated
us to investigate some representatives cases with a larger frequency
cutoff. Here we consider Ohmic and $1/f$ spectrum in Table~\ref{ta3}
(with cutoff). Here the $1/f$-type noise is of interest because it
has been observed in many solid state implementations, especially in
superconducting qubits \cite{ar25,ar26,ar27}.  As seen from
Table~\ref{ta3}, in both cases the two-qubit dephasing can be
greatly suppressed.  Interestingly, although the Ohmic and the $1/f$
cases represent drastically different noise spectrum, in these two
cases the optimized 15-pulse two-qubit DD sequences give similar
performance that is five or six orders of magnitude better than
nested-UDD(3). Clearly then, the improvement afforded by
optimization is sensitive to the cutoff frequency value, but rather
insensitive to the shape of the noise spectrum before the cutoff.
Similar improvements have been observed previously in single qubit
LODD \cite{ar8,ar11}.

\begin{figure}
\scalebox{0.6}[0.6]{\includegraphics{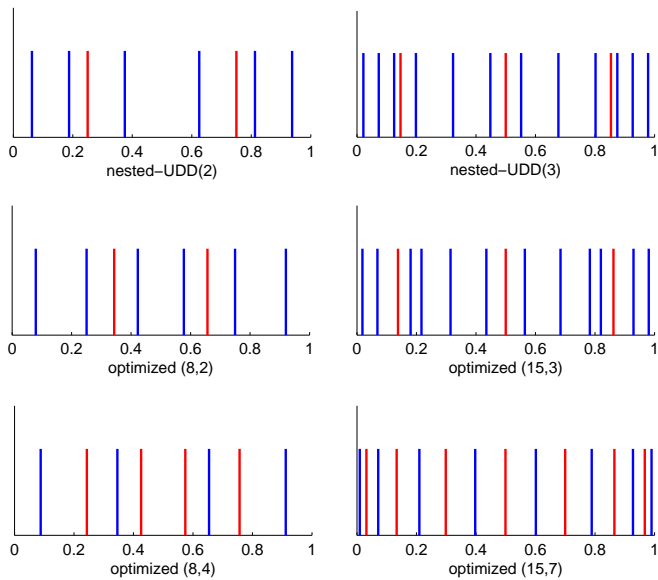}}\caption{Comparsion of specific pulse
locations between nested-UDD and our optimized two-qubit DD sequences.  $S_1=\omega\Theta(5-\omega)$,
$S_2=\omega\Theta(5-\omega)$ and $S_3=\omega\Theta(3-\omega)$.
 The left panel is for $n+m=8$,
and the right panel is for $n+m=15$.   The grey (red) lines stand for timings of the pulses applied to the
second qubit.{\label{fig1}}}
\end{figure}

In Fig.~\ref{fig1} we show
the optimized pulse locations in the case of $S_1=\omega\Theta(5-\omega)$,
$S_2=\omega\Theta(5-\omega)$ and $S_3=\omega\Theta(3-\omega)$.  In the case of $n+m=8$ with $m=2$, the locations of all $\sigma_{x_2}$ pulses in the optimized DD sequence differ remarkably from that in
nested-UDD(2). In the case of $n+m=15$ with $m=3$, the first $\sigma_{x_2}$ pulse is the third pulse in the optimized two-qubit sequence
but the fourth pulse in nested-UDD(3). These observations might be relevant to future experiments.

\subsection{Spectrum with Soft Cutoff}
Noise spectrum with soft cutoff only
gradually decays to zero.
The associated task of decoherence suppression is more challenging.
References~\cite{ar8,ar11,ar12,ar25} showed that for single-qubit dephasing caused by soft-cutoff noise,
UDD can only give a performance
analogous to the Carr-Purcell-Meiboom-Gill (CPMG) sequence \cite{ar28}.  Mathematically,
this can be understood from  the decay exponents $\Gamma_i$ determined by an integral
of a filter function multiplied by a noise spectrum. Because
the noise spectrum is not rapidly vanishing in the high frequency
regime, the behavior of a filter function in the high frequency regime may be important and as a result
the optimization becomes less effective.

Here we study the optimization of DD using various noise spectrum
combinations, involving both super-Ohmic ($\omega^\alpha,\alpha>1$)
spectrum and Lorentzian spectrum with soft cutoff.  The results are
presented in Table~\ref{ta4}. In the first super-Ohmic case with an
exponential soft cutoff, the improvement of optimized DD sequence
over nested-UDD is still magnificent (note that a non-optimized DD sequence
might even speed up the decoherence process~\cite{Shiokawa}).  Indeed, the exponential cutoff
can be regarded as an intermediate case between a truely slow cutoff
(e.g., Lorentzian spectrum) and a hard cutoff studied above. The
second case [$S_1(\omega)=S_2(\omega)=\omega\Theta(1-\omega)$,
$S_3=\frac{0.2}{\omega^2+1}$] should be compared with
Table~\ref{ta2}, with the only difference being that a nonlocal
Ohmic spectrum replaced by a weak nonlocal Lorentzian spectrum.
Consequences of this change in the nonlocal noise spectrum are: (i)
Performance of nested-UDD is greatly reduced, so is the performance
of optimized DD, (ii) With the same value of $n+m$, optimized DD can
only improve the performance over nested-UDD slightly, (iii)
Finally, the performance still improves with the increase of $n+m$,
but very slowly.  Similar observations can be made from the third
case in Table~\ref{ta4}, where the local noise spectrum is assumed
to be Lorentzian.

\begin{table}
\caption{Same as in Table II, but involving noise spectrum without hard cutoff.\label{ta4}}
\begin{ruledtabular}
\begin{tabular}{ccc}
\multicolumn{3}{c}{$S_1=\omega^3e^{-\omega^2},S_2=\omega^3e^{-\omega^2},S_3={\omega}e^{-\omega^2}$}\\
\hline
Pulse combination & Pulse location & Performance \\
\hline
nested-UDD(2) & 3 & $5.31\times10^{-3}$\\
$C_8^4$ & 2,4 & $1.04\times10^{-3}$\\
nested-UDD(3) & 4,8 & $1.44\times10^{-4}$\\
$C_{15}^5$ & 2,5,8 & $5.25\times10^{-9}$\\
\hline
\hline
\multicolumn{3}{c}{$S_1=\omega\Theta(1-\omega),S_2=\omega\Theta(1-\omega),S_3=\frac{0.2}{\omega^2+1}$}\\
\hline
Pulse combination & Pulse location & Performance \\
\hline
nested-UDD(2) & 3 & $4.36\times10^{-3}$\\
$C_8^4$ & 2,4 & $1.67\times10^{-3}$\\
nested-UDD(3) & 4,8 & $1.20\times10^{-3}$\\
$C_{15}^9$ & 2,4,5,7,8 & $4.74\times10^{-4}$\\
\hline
\hline
\multicolumn{3}{c}{$S_1=\frac{0.2}{\omega^2+1},S_2=\frac{0.2}{\omega^2+1},S_3=\omega\Theta(1-\omega)$}\\
\hline
Pulse combination & Pulse location & Performance \\
\hline
nested-UDD(2) & 3 & $2.87\times10^{-2}$\\
$C_8^4$ & 2,4 & $2.08\times10^{-2}$\\
nested-UDD(3) & 4,8 & $1.36\times10^{-2}$\\
$C_{15}^7$ & 2,4,6,8 & $3.96\times10^{-3}$
\end{tabular}
\end{ruledtabular}
\end{table}

\subsection{Asymmetric Local Spectrum}
So far local noise spectrum is assumed to be the same for the two qubits.
In practice the local noise spectrum of individual qubits can be much different. The recent experiment on DD of a two-qubit system represents one excellent example of unbalanced local noise spectrum \cite{ar24}.
In the experiment the dephasing time of a nuclear spin is two orders of magnitude larger than that of
an electron spin. The DD pulses are then applied to the electron spin
only. After a two-flip control sequence, the lifetime of a pseudo-entangled state
in the experiment becomes analogous to that of the nuclear spin, which
implies that control of the nuclear spin is needed to further improve the
entanglement protection.

\begin{figure}
\scalebox{0.6}[0.6]{\includegraphics{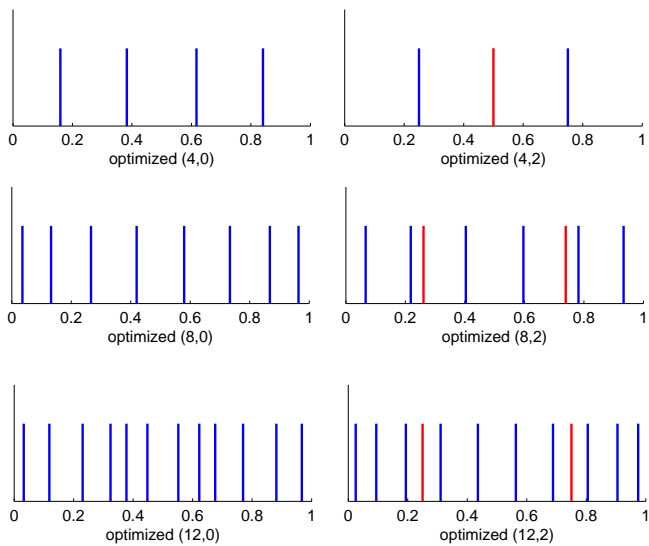}}\caption{Optimized
pulse timings for
$S_1=10\Theta(10-\omega),S_2=0.1\Theta(0.1-\omega),S_3=0.05\Theta(0.05-\omega)$.
The left panels are the results if no pulse is applied to the second qubit. The right
panels depict the optimized DD sequences if two pulses are applied to the second-qubit pulses $(m=2)$. The grey (red) lines stand for timings of the pulses applied to the
second qubit.
From top to bottom, $n+m=4, 8$, and $12$.  Note that in the upper right panel,
the two pulses applied to the second qubit coincide and hence cancel each other.
{\label{fig2}}}
\end{figure}

To mimic an environment like the experimental setup in
Ref.~\cite{ar24}, we consider an environment with strongly
asymmetric local noise.  For convenience the noise spectrum is
assumed to be a constant subject to a hard cutoff, i.e.,
$S_1(\omega)=10\Theta(10-\omega)$,
$S_2(\omega)=0.1\Theta(0.1-\omega)$, and
$S_3(\omega)=0.05\Theta(0.05-\Omega)$. For these noise parameters
the dimensionless dephasing rate of the first qubit
is about two orders of magnitude larger than that of
the second qubit.  The optimization results are shown in
Table~\ref{ta5}. At least two interesting observations can be made.
First, in the case of $n+m=4$, assigning some control pulses to the
second qubit may bring down the performance. Hence, if the local
noise spectrum is strongly unbalanced and if $n+m$ is highly
limited,  then it is beneficial to control only one of the two
qubits. This is consistent with the experimental implementation in
Ref.~\cite{ar24}. Second, when the minimized total error approaches
the error caused by the free decay of the second qubit, solely
applying $\sigma_{x_1}$ pulses  can no longer improve the DD
performance. As shown in Table~\ref{ta5}, the DD performance without
any $\sigma_{x_2}$ pulses is bounded by about $10^{-2}$.  However,
if now two $\sigma_{x_2}$ pulses are applied,  then the DD performance can be improved by
many orders of magnitude (e.g., in the 12-pulse case). Note also
that for fixed $n+m$, increasing $m$ too much undermines the
performance. Again, this is because the number of pulses applied to the two qubits should be balanced so that
errors of the two qubits are suppressed to the same level. Less pulses should be applied to the second qubit because it is more weakly coupled to its environment. As such,  the number of $\sigma_{x_1}$ pulses
($n$) should always dominate. In Fig.~\ref{fig2} we display the
optimized pulse locations in several cases.  The left panels of
Fig.~\ref{fig2} represent optimized pulse locations if $m=0$.  The
right panels show the optimized pulse locations
 if $m=2$. Interestingly, in the upper right panel, it is found that
for $n+m=4$, the two pulses for the second qubit should be applied at the same time,
thus canceling themselves.
Our optimization procedure therefore requests a null action on the second qubit if $n+m$ is small.
For the case of optimized (8,2) shown in Fig.~2, the pulse combination is the same as in nested-UDD(2), but with optimized pulse timings. We have also compared the optimized (12,2) case in Fig.~2 with an asymmetric nested-UDD of total 11 pulses,  namely,  UDD(3) on the first qubit in the inner layer and UDD(2) on the second qubit in the outer layer. The performance of this asymmetric nested-UDD is only $0.517$, still far from what is achieved here (only at the cost of one additional pulse in the middle interval of the inner layer).

\begin{table}
\caption{Same as in Table II, but for an example where the local environments of the two qubits are highly
 unbalanced.\label{ta5}}
\begin{ruledtabular}
\begin{tabular}{ccc}
\multicolumn{3}{c}{$S_1=10\Theta(10-\omega),S_2=0.1\Theta(0.1-\omega),S_3=0.05\Theta(0.05-\omega)$}\\
\hline
Pulse combination & Pulse location & Performance \\
\hline
$C_4^0$ & no pulse & $1.30$\\
$C_4^2$ & 2 & $2.00$\\
$C_8^0$ & no pulse & $2.00\times10^{-2}$\\
$C_8^2$ & 3 & $7.64\times10^{-3}$\\
$C_8^4$ & 1,3 & $1.30$\\
$C_8^6$ & 1,2,4 & $2.00$\\
$C_{12}^0$ & no pulse & $1.99\times10^{-2}$\\
$C_{12}^2$ & 4 & $1.57\times10^{-7}$\\
$C_{12}^4$ & 3,5 & $6.25\times10^{-6}$\\
$C_{12}^6$ & 2,4,6 & $7.45\times10^{-3}$\\
$C_{12}^8$ & 1,3,4,6 & $1.29$
\end{tabular}
\end{ruledtabular}
\end{table}

\subsection{Limitations on Optimized DD}
As in single-qubit DD, we would naturally attempt to increase the total pulse number $n+m$
to achieve better and better DD performance. However, the optimization procedure faces
much difficulty when $n+m$ increases considerably.  First, the optimization
algorithm becomes much slower in each run and may fail to give a converged result. For example,
for $n+m=24$, the converged pulse
locations might be inappropriate, i.e., violating the initial ordering of the pulse locations.
Worse still, for $n+m=24$, we have to run the
algorithm $924$ times for the single combination $C_{24}^{12}$ (12 pulses for the second qubit).  Hence computationally it is prohibitively expensive if hundreds of pulses need to be optimized.
Second, as the pulse number increases, the optimized result becomes
more sensitive to the initial guess.  Mathematically it is also a challenging
question to find the global minimum in a high-dimensional
parameter space. For $n+m=24$, we have considered two initial guesses, namely, an
equally spaced DD sequence and a nested-UDD(4) sequence. The associated results
are shown in Table~\ref{ta6}. It is seen that for the same noise spectrum,
the performance is only slightly better than that for $n+m=15$. It is hence possible
that our result as a locally optimized result is still far away from the globally optimized result.
More effective optimization algorithms together with more initial guesses may be the ultimate solution.

\begin{table}
\caption{Comparison between optimized two-qubit DD with $n+m=24$ and nested-UDD(4). The meaning of the table columns are the same as in Table II.
Noise spectrum is similar to that considered in Tables II and III.  \label{ta6}}
\begin{ruledtabular}
\begin{tabular}{ccc}
\multicolumn{3}{c}{$S_1=\omega\Theta(1-\omega),S_2=\omega\Theta(1-\omega),S_3=0.5\omega\Theta(0.5-\omega)$}\\
\hline
Pulse combination & Pulse location & Performance \\
\hline
nested-UDD(4) & 5,10 & $5.21\times10^{-9}$\\
$C_{24}^4$ & 2,9 & $2.81\times10^{-10}$\\
$C_{24}^8$ & 1,3,5,11 & $3.31\times10^{-11}$\\
$C_{24}^{12}$ & 2,4,7,8,10,12 & $2.34\times10^{-11}$\\
\hline
\hline
\multicolumn{3}{c}{$S_1=\omega\Theta(5-\omega),S_2=\omega\Theta(5-\omega),S_3=\omega\Theta(3-\omega)$}\\
\hline
Pulse combination & Pulse location & Performance \\
\hline
nested-UDD(4) & 5,10 & $3.31\times10^{-2}$\\
$C_{24}^4$ & 3,9 & $1.42\times10^{-3}$\\
$C_{24}^8$ & 1,3,5,10 & $1.51\times10^{-7}$\\
$C_{24}^{12}$ & 2,4,6,9,11,12 & $1.35\times10^{-7}$
\end{tabular}
\end{ruledtabular}
\end{table}

It should be also noted that the total pulse number is physically limited. Realistic
pulses are imperfect and an ideal instantaneous $\pi$ rotation is
never possible \cite{ar29}. As a result, the pulse-to-pulse errors
may accumulate with increase of the pulse number~\cite{ar16,ar30} and hence
more pulses do not necessarily lead to better performance.
Furthermore, constraint on the minimal pulse interval might place another
intrinsic limit on the pulse number we could consider \cite{ar31}. For
fixed total evolution time $t$, such a constraint imposes an
upper bound for the pulse number. For a varying $t$, the
optimized performance in practice will be connected with the available pulsing rate
as well as the spectral bandwidth \cite{ar32,ar33}.

\section{CONCLUSIONS}
Using a pure dephasing model incorporating both local and nonlocal
Gaussian noise, we have studied how the DD protection of two-qubit states
can be better achieved by optimizing the pulse locations as well as the partition of the pulse
numbers allocated to each qubit.  Compared with nested-UDD as a general-purpose DD scheme for two-qubit systems,
our optimization procedure may improve the performance by many orders of magnitude, using only a few tens of instantaneous $\pi$ pulses.
This makes it possible to use much less pulses to obtain a given
fidelity of decoherence control in two-qubit systems.
The price of this performance gain is the required knowledge of the noise spectrum.  In addition, it is also seen that
if the noise spectrum decays to zero slowly (e.g.,
a Lorentzian shape), then even our optimized DD may not perform very well and it is only slightly better than nested-UDD.

The results here also help to gain more insights into the issue of DD protection in two-qubit or multi-qubit systems.
The importance of fighting against the nonlocal noise in two-qubit decoherence control (or more generally, the importance of taking into account the impact of the dynamics of one qubit on the environment of the other qubit) becomes clearer.  Without the nonlocal
noise, our optimized DD sequence simply degenerates into two single-qubit optimized DD
sequences. In the presence of nonlocal noise, the optimized two-qubit DD sequence is quite
different from two optimized single-qubit DD sequences, because
the pulse numbers and pulse locations for each qubit should be adjusted (in a somewhat subtle manner)
based on both nonlocal and local noise.
If the
local noise spectrum of the two qubits is highly unbalanced, then most pulses should be applied to
one of the two qubits to defeat its rapid dephasing.  It is these specific strategies, which are not exploited
by the rather universal nested-UDD scheme,
that makes optimization possible.  Note also that in nested-UDD schemes, the number of control pulses or the control order
may be different for different control layers. But as seen from one optimization result in Sec.~IV-C, the performance of
such type of asymmetric-nested-UDD is still far from optimized.

There have been wide experimental interests in both fundamental aspects of decoherence and the DD approach to decoherence suppression.  Two types of experiments may be motivated by this work. First, for two-qubit systems with known noise spectrum (local and nonlocal), it is of immediate interest to apply optimized DD sequences to extend the lifetime of entangled states, hopefully with better efficiency. Second, one may experimentally study the nonlocal noise of a two-qubit system by use of single-qubit DD. That is, if two single-qubit DD sequences are applied on top of each other, then (assuming that the local noise of the two qubits are independent) only the nonlocal noise is not suppressed and hence its impact on two-qubit dephasing may be directly observed. Such type of experiments are of importance to the
design of optimized two-qubit DD sequences and to the understanding of decoherence mechanism in two-qubit systems embedded in a solid-state environment.

\begin{acknowledgments}
This collaborative work is made possible by the International Collaboration Fund from the Faculty of Science,
National University Singapore, and by
NUS ``YIA" (R-144-000-195-101). Z.R.Xi was supported by the National Nature Science
Foundation of China under Grant Nos. 61074051 and
60821091.  J.G. acknowledges interesting discussions with Ya Wang, Pu Huang, Dawei Lu, and Jiangfeng Du from USTC.

\end{acknowledgments}

\end{document}